\documentclass[12pt,cleanfoot]{asme2ej}

\usepackage{graphicx}
\usepackage{fancyhdr}
\usepackage{setspace}
\usepackage{helvet}
\usepackage[hyphens]{url}

\usepackage{hyperref}   % to set up hyperlinks
\hypersetup{
	colorlinks=true,
	linkcolor=blue,
	citecolor=blue,
	urlcolor=blue,
}
\usepackage[square,numbers,sort&compress]{natbib}

\usepackage{amsmath,amssymb,amsfonts}
\usepackage{algorithmic}
\usepackage{textcomp}
\usepackage{upgreek}
\usepackage{soul,xcolor}
\soulregister{\cite}7
\soulregister{\ref}7
\usepackage{epstopdf}
\graphicspath{{./fig/}}
\usepackage{caption}
\title{Thermal spreading resistance of GaN HEMTs with heat source heating studied by hybrid Monte Carlo-diffusion simulations}
\author{Han-Ling Li
    \affiliation{
	Key laboratory of thermal science and\\
    power engineering of Education of Ministry\\
	Department of Mechanical Engineering\\
	Tsinghua University\\
	Beijing, 10084\\
    China\\
    Email: lihanling1994@163.com
    }	
}

\author{Yang Shen
    \affiliation{
	Key laboratory of thermal science and\\ 
    power engineering of Education of Ministry\\
	Department of Mechanical Engineering\\
	Tsinghua University\\
	Beijing, 10084\\
    China\\
    Email: sheny21@mails.tsinghua.edu.cn
    }	
}

\author{Yu-Chao Hua
    \affiliation{
    LTEN laboratory\\
	Polytech Nantes\\
	University of Nantes\\
	Nantes, UMR6607, F-44000\\
    France\\
    Email: huayuchao19@163.com
    }	
}

\author{S.L. Sobolev
    \affiliation{
    Institute of Problems of Chemical Physics\\
	Academy of Sciences of Russia\\
	Chernogolovka, Moscow Region, 142432\\
    Russia\\
    Samara State Technical University\\
    ul. Molodogvardeiskaya 244, Samara, 443100\\
    Russia\\
    Email: sobolev@icp.ac.ru
    }	
}

\author{Bing-Yang Cao\thanks{Corresponding author}
    \affiliation{
	Key laboratory of thermal science and power engineering of Education of Ministry\\
	Department of Mechanical Engineering\\
	Tsinghua University\\
	Beijing, 10084\\
    China\\
    Email: caoby@mail.tsinghua.edu.cn
    }	
}

\begin{document}

\maketitle

\doublespacing

\vspace{1cm}
\begin{abstract}
    {\it
    Exact assessment of thermal spreading resistance is of great importance to the thermal management of electronic devices, especially when completely considering the heat conduction process from the nanoscale heat source to the macroscopic scale heat sink. The existing simulation methods are either based on convectional Fourier's law or limited to small system sizes, making it difficult to accurately and efficiently study the cross-scale heat transfer. In this paper, a hybrid phonon Monte Carlo-diffusion method that couples phonon Monte Carlo (MC) method with Fourier's law by dividing the computational domain is adopted to analyze thermal spreading resistance in ballistic-diffusive regime. Compared with phonon MC simulation, the junction temperature of the hybrid method has the same precision, while the time costs could be reduced up to 2 orders of magnitude at most. Furthermore, the simulation results indicate that the heating scheme has a remarkable impact on phonon transport. The thermal resistance of the heat source (HS) scheme can be larger than that of the heat flux (HF) scheme, which is opposite from the prediction of Fourier's law. In the HS scheme, the enhanced phonon-boundary scattering counteracts the broadening of the heat source, leading to a stronger ballistic effect as the heat source thickness decreases. The conclusion is verified by a one-dimensional thermal resistance model. This work has opened up an opportunity for the fast and extensive thermal modeling of cross-scale heat transfer in electronic devices and highlighted the influence of heating schemes.

    }
\end{abstract}
	
\section{Introduction}
\label{sec:introduction}
\begin{figure}[htbp]
    \centering
    \includegraphics[width=80mm]{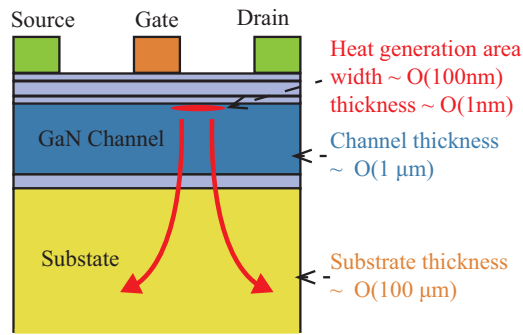}
    \caption{Schematic of GaN HEMTs}
    \label{fig: Schematic of GaN HEMTs}
\end{figure}

Owing to its advantages in high breakdown voltage and large bandgap, the gallium nitride (GaN) high-electron-mobility transistor (HEMT) is an excellent choice for high-voltage and high-frequency power electronic devices \cite{MishraParikh-1043,HiramaKasu-1044}. The power improvement in GaN HEMTs inevitably leads to huge power density ($>10\;{\rm kW/cm^2}$) \cite{Bar-CohenAlbrecht-1045} and elevated junction temperature, which results in significant thermal bottlenecks of these devices \cite{ZanoniMeneghesso-1055,MeneghessoVerzellesi-1054, Stocco-1048,Bagnall-786,PainePolmanter-1056}. The typical structure of GaN HEMTs is shown in Fig. \ref{fig: Schematic of GaN HEMTs}, and its basic form is made of multilayer films. During operation, heat is primarily generated in the channel layer under the drain side edge of the gate \cite{SridharanVenkatachalam-830}, the width of the heat generation region is on the order of 100 nm, while the thickness is on the order of 1 nm \cite{DonmezerGraham-832}. The thickness of the channel layer is usually 1 - 3 $\upmu \rm m$, whereas the thickness of the substrate is larger than 100 $\upmu \rm m$. It is worth noting that the length and width of the channel layer can reach the order of 1 mm. In this way, the generated heat will spread from a small area to a much larger substrate, as the arrows in Fig. \ref{fig: Schematic of GaN HEMTs} show. This phenomenon produces a multidimensional temperature field and a conspicuous near junction resistance, which is also known as thermal spreading resistance \cite{Kennedy-1057}. In addition, the thicknesses of the heat generation region, channel layer and substrate vary several orders of magnitude, resulting in a cross-scale heat conduction process. Since it is a pivotal problem in the heat transfer of HEMTs, accurate and efficient evaluation of thermal spreading resistance is critical to predict the junction temperature and assess the reliability, as well as developing innovative thermal designs \cite{WonCho-1051,MooreShi-360,WonCho-1053,GarimellaPersoons-492}.    

Based on Fourier's heat conduction law \cite{Fourier-982}, a large amount of work has been reported on the calculation of thermal spreading resistance. Krane \cite{Krane-906} established analytical models for thermal spreading resistance under the conditions of a concentric heat source on rectangle channels. Then, Muzychka et al. \cite{MuzychkaCulham-828} studied the cases of eccentric and discrete heat sources, and Darwish et al. \cite{DarwishBayba-396} investigated the effect of the substrate. Moreover, the model has been expanded to consider the interfacial thermal resistance \cite{MuzychkaBagnall-421}, the anisotropy of thermal conductivity \cite{GholamiBahrami-829}, the arbitrary number of layers \cite{BagnallMuzychka-420} and the temperature-dependent thermal conductivity \cite{DarwishBayba-491}. However, directly adopting Fourier's law to heat transfer in micro and nanoscale electronic devices will cause some errors \cite{PopSinha-40,CahillBraun-359,WarzohaWilson-1002}. Mean-free-paths (MFPs) of phonons (the major heat carriers in semiconductors) for GaN are approximately 100 nm to 10 $\upmu$m \cite{MaWang-1059,FreedmanLeach-273,ZiadeYang-1058}, which are comparable to the width of heat generation region and thickness of the channel layer in GaN HEMTs. As a result, phonon ballistic transport emerges \cite{Chen-3, MaznevCuffe-64}, and ballistic-diffusive heat conduction raises the junction temperature to be higher than the prediction of Fourier's law \cite{Chen-495,SchleehMateos-38,HuaCao-9,LiCao-538}.

Efforts have been devoted to exploring the characteristics of ballistic-diffusive heat conduction and accounting for phonon ballistic transport in the thermal analyses of electronic devices. Cao and Li \cite{CaoLi-507,LiCao-506} studied the effective thermal conductivity of nanostructures by nonequilibrium molecular dynamics (MD) simulations and found that the results of the uniform heat source (HS) scheme are much lower than those of the temperature difference scheme. Using phonon Monte Carlo (MC) simulations, Hua and Cao \cite{HuaCao-70} obtained similar results and established a theoretical model for the effective thermal conductivity in the internal HS scheme. Then, it is claimed that the models of the effective thermal conductivities can be unified as a function of the Knudsen number (ratio of phonon MFP to the characteristic length) and the shape factor \cite{HuaCao-67}. Recently, Hua et al. \cite{HuaLi-661} investigated the thermal spreading resistance in ballistic-diffusive regime under the heat flux (HF) heating scheme and demonstrated that there are two ballistic effects affecting heat transfer: one is related to the cross-plane Knudsen number, and the other is related to the lateral Knudsen number. These studies suggest that the geometric size and heating scheme play a significant role in ballistic-diffusive heat conduction. However, subject to computational complexity and costs, MD or phonon MC is limited to small systems and cannot simulate the cross-scale heat transfer process in real electronic devices \cite{BaoChen-548}.
The hybrid method that combines the accuracy of the micro and nanoscale simulation techniques and the simplicity of Fourier's law is a promising way to overcome this problem \cite{ChoiGraham-984}. By solving phonon Boltzmann transport equation (BTE) in a small domain of the device channel and utilizing Fourier's law in the rest domain, Donmezer and Graham \cite{DonmezerGraham-832} found that the hotspot temperature is higher when the ballistic-diffusive transport effect is considered. Hao et al. \cite{HaoZhao-993} combined phonon MC and Fourier's law in a similar way to simulate two-dimensional (2D) GaN HEMTs, then the method was applied to more complex structures \cite{HaoZhao-997, HaoZhao-1016}. Chatterjee et al. \cite{ChatterjeeDundar-823} proposed a phonon BTE - Fourier coupled thermal modeling to study the interplay of heat concentration and ballistic effect. These reports successfully simulate the heat transfer of GaN HEMTs in particular sizes with the HS schemes, but detailed research of the thermal spreading resistance in systems with a relatively large size is still lacking. Moreover, the influence of heating schemes on the calculation of thermal spreading resistance remains unclear.

In this paper, a hybrid phonon MC-diffusion method is employed to study the thermal spreading resistance in ballistic-diffusive regime. By dividing the whole system and coupling phonon MC simulation with Fourier's law, the hybrid method is capable of characterizing the heat transfer process as accurately as phonon MC simulation while greatly scaling down the computational time, facilitating the fast simulation of thermal spreading resistance over a comprehensive size range. The heating scheme is found to have a remarkable impact on phonon transport, and the ballistic effect in the HS scheme can be much stronger than that in the HF scheme as the thickness of the heat generation region declines. Parametric investigations and theoretical analyses claim that this can be interpreted by the variation in the heat source spatial distribution and phonon boundary scattering.

\section{Methods}
\begin{figure}[htbp]
    \centering
    \includegraphics[width=80mm]{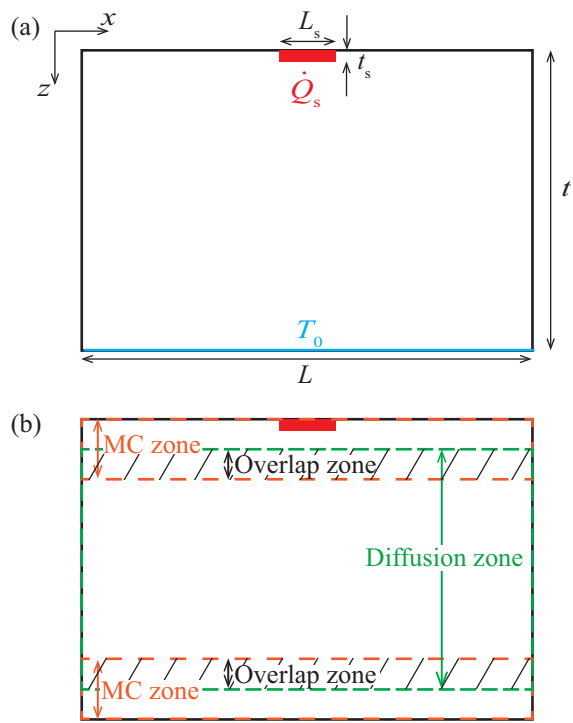}
    \caption{Schematic of the research object: (a) the 2D simplified model and (b) domain dividing scheme.}
    \label{fig: Schematic of the research object}
\end{figure}

Fig. \ref{fig: Schematic of the research object} (a) shows the research object of this work, which is a representative 2D simplified model for the heat dissipation process in GaN HEMTs. The substrate material is assumed to be the same as the channel material, that is, GaN, and the interface between the channel and the substrate is ignored. Compared to the system discussed in \cite{HuaLi-661}, the heat generation is modeled as an internal heat source in a rectangle region of size $L_{\rm s} \times t_{\rm s}$ with a power dissipation density of $\dot{Q}_{\rm s}$, instead of boundary heat flux. The HS scheme agrees better with the results of electrothermal simulations \cite{SridharanVenkatachalam-830,HaoZhao-993,ChatterjeeDundar-823}; thus, it is expected to simulate the heat transfer process more realistically than the HF scheme. The boundary conditions are consistent with \cite{HuaLi-661}, namely, fixed temperature $T_0$ at the bottom boundary, adiabatic at the top boundary, and periodic at the lateral boundaries. In our calculations, the material property of GaN at 300 K is adopted, the heating power is fixed at $P=L_{\rm s}t_{\rm s}\dot{Q}_{\rm s}=0.2\;{\rm W}$, the ambient temperature is set as $T_0=300\;{\rm K}$. Structures with a variety of sizes will be investigated, and the corresponding values of $L$, $t$, $L_{\rm s}$ and $t_{\rm s}$ will be given in Sec. \ref{sec: Results and discussions}.

The hybrid MC-diffusion method developed in \cite{LiHua-469} is employed to characterize ballistic-diffusive heat conduction, as shown in Fig. \ref{fig: Schematic of the research object} (b). Based on the idea that ballistic transport mainly affects the regions adjacent to the phonon source and boundaries when the system is considerably large, the whole computational domain is divided into three zones: the top MC zone that covers the heat generation region and top boundary, the bottom MC zone that covers the bottom boundary, and the middle diffusion zone. There are overlap zones between the MC and diffusion zones to transfer information and check convergence. The thicknesses of the top MC zone, bottom MC zone and overlap zone are denoted as $t_{\rm topMC}$, $t_{\rm botMC}$ and $t_{\rm o}$. Compared with the hybrid BTE-Fourier methods in \cite{HaoZhao-993,HaoZhao-997, HaoZhao-1016,ChatterjeeDundar-823}, there is an additional bottom MC zone, of which the necessity will be explained in Sec. \ref{sec: validation}

\begin{figure}[htbp]
    \centering
    \includegraphics[width=80mm]{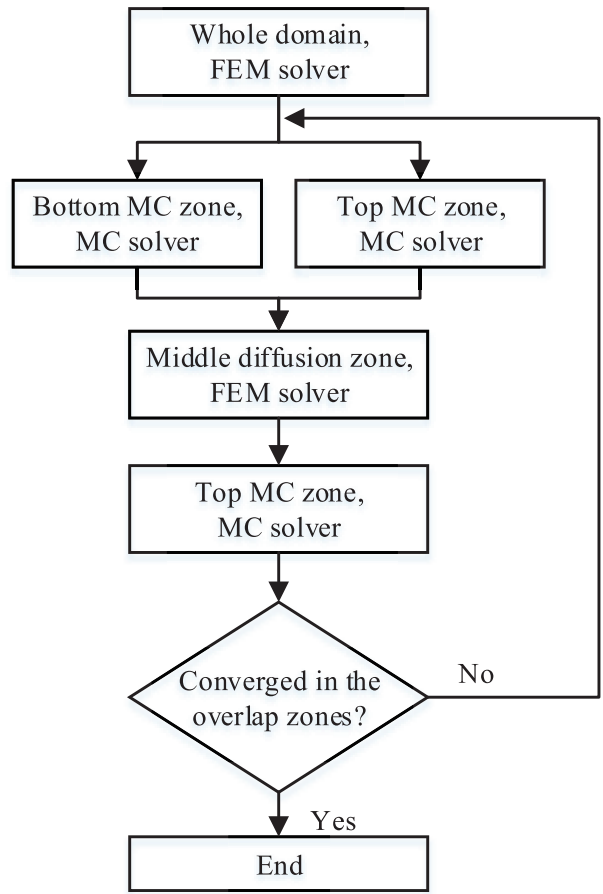}
    \caption{Flowchart of the hybrid MC-diffusion method}
    \label{fig: Flowchart of the hybrid method}
\end{figure}

Phono MC simulation and the finite element method (FEM) based on Fourier's law are coupled by an alternating technique in the hybrid method. The flowchart of the hybrid method is illustrated in Fig. \ref{fig: Flowchart of the hybrid method}, which has the following procedures: \textbf{(\romannumeral1) Initialization:} Use the FEM to obtain the diffusive solution of the problem. \textbf{(\romannumeral2) Phonon MC simulation:} Use the heat flux obtained by FEM as boundary conditions at the interfaces from the MC zones to the diffusion zone and run phonon MC simulation. \textbf{(\romannumeral3) Middle diffusion solution:} Alternate the boundary temperature and heat flux at the interfaces from the diffusion zone to the MC zones by the results of MC simulations and run FEM. \textbf{(\romannumeral4) Hot spot temperature refinement:} Run the MC simulation again in the top MC domain with the FEM-updated boundary temperature. \textbf{(\romannumeral5) Convergence judgment:} If the temperature and heat flux distribution of phonon MC and FEM converge to equality in the overlap zones, the calculation ends; otherwise it proceeds to (\romannumeral2) and repeats. 

In this paper, the gray-body approximation that assumes all phonons have the same MFP (denoted as $\Lambda$) is used to simplify the problem. With a proper and representative choice of the MFP, the results of the gray-body approximation are in good agreement with the results on account of phonon dispersion as well as experimental measurements \cite{LiShiomi-838}. In-depth discussions about the effect of phonon dispersion on thermal spreading resistance have been carried out elsewhere \cite{shen2022spectral}.
Noting that the size of the region affected by the phonon source or phonon absorbing boundary is around one phonon MFP \cite{Chen-495}, the sizes of the MC zones and overlap zones are set as $t_{\rm topMC}=\max(2.0\Lambda,t_{\rm s}+1.5\Lambda)$, $t_{\rm botMC}=2.0\Lambda$, $t_{\rm o}=1.0\Lambda$ \cite{LiHua-469}. Such settings obtain a pleasant balance between the calculation accuracy and efficiency. The number of simulated phonon bundles is $10^7$, which is large enough to suppress the random error originating from the statistical nature of MC simulation. Details of phonon MC simulation and the hybrid MC-diffusion method can be found in \cite{HuaCao-9} and \cite{LiHua-469}, respectively.

The material difference of the channel layer and the substrate will result in different temperature gradient, which can be taken into consideration by changing phonon properties during the calculation. Thermal boundary resistance (TBR) \cite{PohlSwartz-392} between the channel layer and substrate is another important factor in the heat dissipation of GaN HEMTs. TBR brings about a temperature discontinuity at the material interface and increases the junction temperature \cite{GarciaIniguezdelatorre-416}. In phonon MC simulation, TBR is usually treated as the transmission and reflection interface condition \cite{HuaCao-499,RanGuo-638}, in which the phonon transmissivity needs to be given in advance by models or experiments. However, the existence of material change and TBR do not affect the basic idea of the hybrid method or the the influence of heating schemes on thermal spreading resistance. To focus on what we are concerned with, material property difference and TBR are neglected in this paper.

\section{Results and discussions}
\label{sec: Results and discussions}
\subsection{Validation of the hybrid MC-diffusion method}
\label{sec: validation}
\begin{figure}[h]
    \centering
    \begin{minipage}{53mm}
        \centering
        \includegraphics[width=1.0\linewidth]{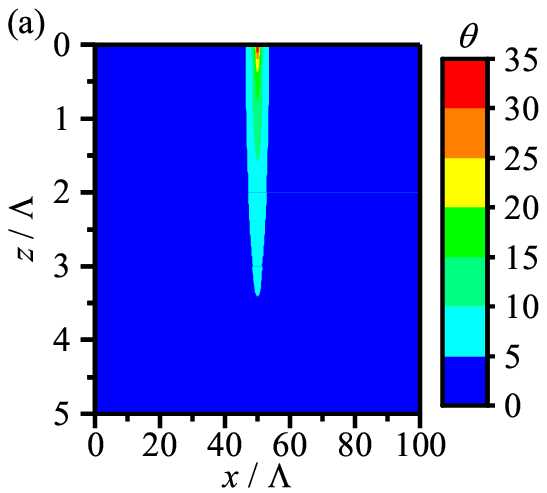}
    \end{minipage}
    \begin{minipage}{53mm}
        \centering
        \includegraphics[width=1.0\linewidth]{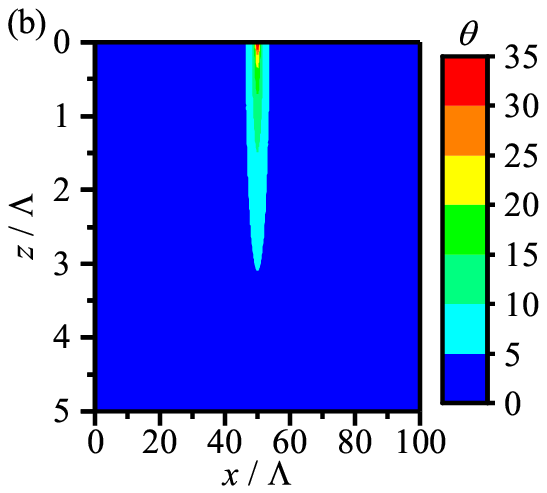}
    \end{minipage}
    \begin{minipage}{53mm}
        \centering
        \includegraphics[width=1.0\linewidth]{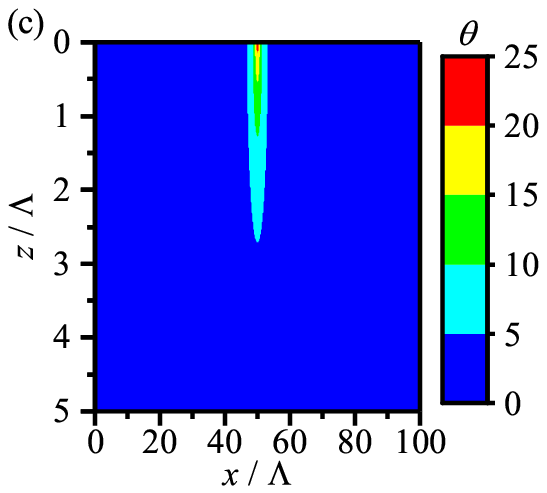}
    \end{minipage}
    \caption{Dimensionless temperature distributions calculated by (a) hybrid MC-diffusion method (b) MC simulation and (c) FEM. The sizes are $Kn_L=0.01$, $Kn_t=0.2$, $Kn_{L,{\rm s}}=1$, $Kn_{t,{\rm s}}=10$.}
    \label{fig: theta distributions}
\end{figure}

At first, the performance of the hybrid MC-diffusion method when used to thermal spreading problems is examined. According to the typical size of GaN HEMT, the four geometry parameters in the test case are set as: $Kn_L=0.01$, $Kn_t=0.2$, $Kn_{L,{\rm s}}=1$, $Kn_{t,{\rm s}}=10$, in which $Kn$ denotes the Knudsen number. The dimensionless temperature distributions predicted by the hybrid MC-diffusion method are shown in Fig. \ref{fig: theta distributions} (a). As a comparison, the results calculated by the MC simulation and FEM are depicted in Fig. \ref{fig: theta distributions} (b) and Fig. \ref{fig: theta distributions} (c), respectively. The dimensionless temperature is defined as $\theta=(T-T_0)/(\dot{Q}_{\rm s} L_{\rm s} t_{\rm s} R_{\rm 1D,0})$ where $R_{\rm 1D,0}=t/(Lk_{\rm bulk})$ is the one-dimensional (1D) diffusive thermal resistance in the HF scheme \cite{HuaLi-661}. The temperature distribution of the hybrid MC-diffusion method is in reasonable agreement with that of the MC simulation; they both predict longer and narrower hot areas and higher peak temperatures than the FEM results, indicating that the hybrid MC-diffusion method successfully characterizes phonon ballistic transport.

\begin{figure}[htbp]
	\centering
    \includegraphics[width=80mm,height=70mm]{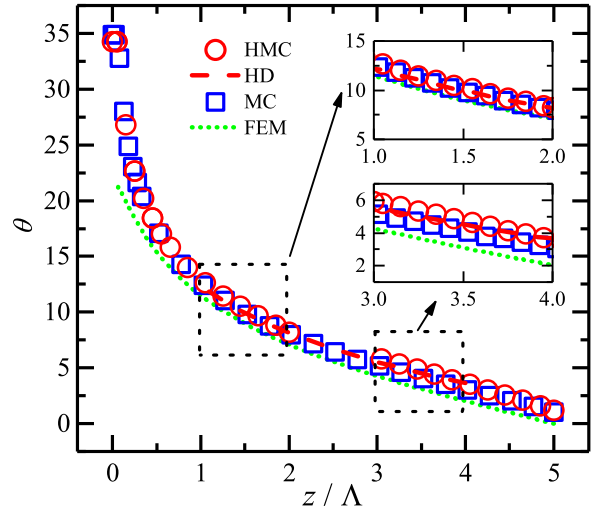}
    \caption{Dimensionless temperature at the vertical symmetry line ($Kn_L=0.01$, $Kn_t=0.2$, $Kn_{L,{\rm s}}=1$, $Kn_{t,{\rm s}}=10$). ``HMC'' and ``HD'' denote the hybrid MC and hybrid diffusion, respectively. The insets show the enlarged figures of the top and bottom overlap zones with more nodes displayed.}
    \label{fig: theta xmid}
\end{figure}

To better reveal the accuracy of the hybrid MC-diffusion method, the dimensionless temperature along the $z$ direction at the vertical symmetry line is drawn in Fig. \ref{fig: theta xmid}. The temperature convergence of the hybrid MC (HMC) and hybrid diffusion (HD) at the two overlap zones is clearly demonstrated by the insets, and the hybrid method has a great consistency with the MC simulation. For the concerned junction temperature at $z=0$, the result of the hybrid MC-diffusion method is $\theta_{\rm max, hybrid}=34.2$, while the value of the MC simulation is $\theta_{\rm max, MC}=34.9$, and their relative error is only about $2\%$. Compared to the   FEM results, the temperature profile of the hybrid MC-diffusion method is always higher, even in the bottom MC zone ($3\Lambda \le z \le 5\Lambda$). Although the bottom MC zone has a distance of more than $3\Lambda$ from the heat generation region, since the system thickness is only a few phonon MFPs, phonon-boundary scattering at the bottom boundary has played a remarkable role, leading to a slight but obvious boundary temperature jump \cite{HuaCao-253}. Therefore, Fourier's law fails at the regions near the bottom boundary, and it is necessary to introduce the bottom MC zone in the hybrid method to capture such a boundary nonequilibrium phenomenon.

\begin{figure}[htbp]
	\centering
    \includegraphics[width=80mm,height=70mm]{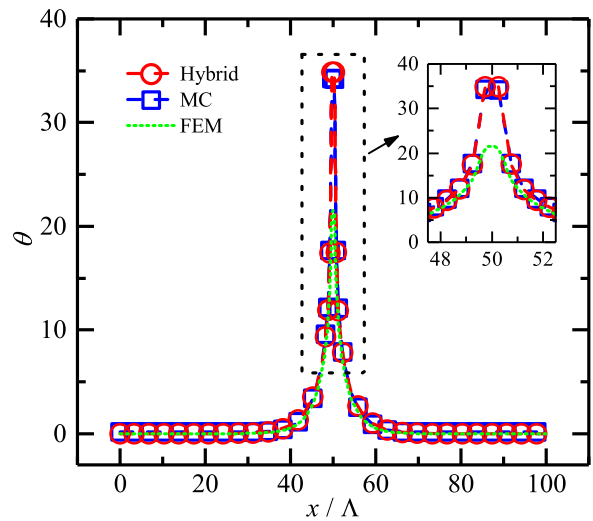}
    \caption{Dimensionless temperature at the top boundary ($Kn_L=0.01$, $Kn_t=0.2$, $Kn_{L,{\rm s}}=1$, $Kn_{t,{\rm s}}=10$)}
    \label{fig: theta at z0}
\end{figure}

In addition, the temperature distribution at the top of the system, which could be experimentally measured in a real HEMT \cite{ChatterjeeDundar-823}, is drawn in Fig. \ref{fig: theta at z0}. As expected, the hybrid MC-diffusion method agrees well with the MC simulation and predicts a considerably higher temperature than FEM around the heat generation region. Away from the heat generation region, the difference between hybrid the MC-diffusion method and FEM fades out, and the temperature can be calculated by Fourier's law. It is worth noting that most phonon movement and scattering occur near the heat generation region, to which the vast majority of computing resources are devoted under the energy-based variance-reduced MC framework \cite{PeraudHadjiconstantinou-50}. The area other than the heat generation region contributes little to the computational cost, so the current hybrid method lets the length of the top MC zone be the same as the system length to simplify the iteration process.

\begin{figure}
    \centering
    \begin{minipage}{80mm}
        \centering
        \includegraphics[width=1.0\linewidth,height=70mm]{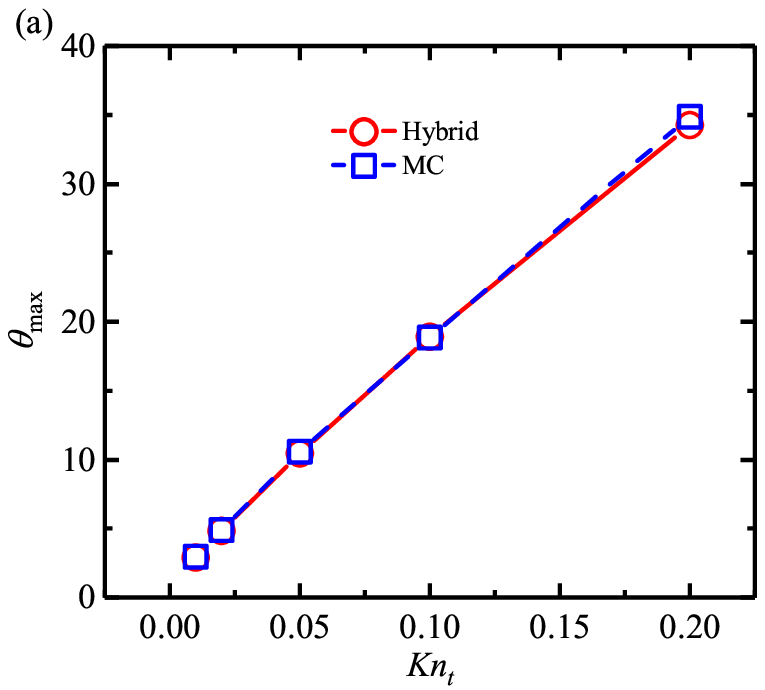}
    \end{minipage}
    \begin{minipage}{80mm}
        \centering
        \includegraphics[width=1.0\linewidth,height=70mm]{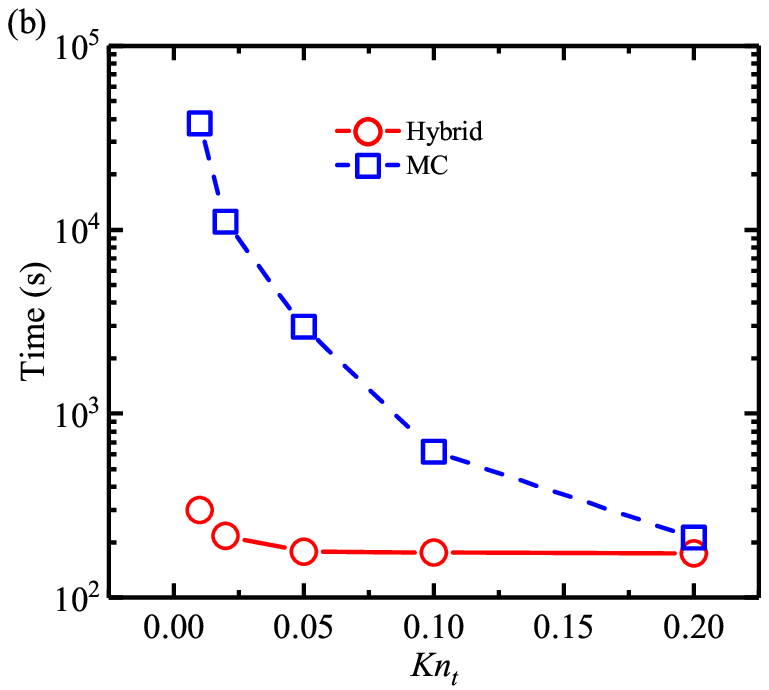}
    \end{minipage}
    \caption{(a) Dimensionless junction temperature varying with $Kn_t$; (b) computational time varying with $Kn_t$.}
    \label{fig: performance of hybrid method}
\end{figure}

More geometric structures are calculated to test the accuracy and efficiency of the hybrid MC-diffusion method. Fig. \ref{fig: performance of hybrid method} illustrates the dimensionless junction temperature and computational time of the hybrid MC-diffusion method varying with the system thickness, while the other three geometry parameters are kept the same. As a benchmark, the results of the MC simulation are also given. It can be seen that the hybrid method has a much better efficiency compared to the MC simulation without causing significant deviation on $\theta_{\rm max}$, and the computational time can be mostly reduced by up to 2 orders of magnitude, from tens of thousands seconds to hundreds of seconds. MC simulation could be an efficient tool when $Kn_t$ is relatively large, as its computation time greatly decreases with $Kn_t$. For the hybrid MC-diffusion method, since the main time-consuming process is the MC simulation in the MC zones, the computation time is expected to be a constant as the sizes of the MC zone are fixed, which is verified by the nearly horizontal red line in Fig. \ref{fig: performance of hybrid method} (b). The small oscillations are due to the different number of iterations required for convergence. It is recommended that for $Kn_t > 0.2$, there is no need to use the hybrid MC-diffusion method because the MC simulation is fast enough.  

\subsection{Thermal spreading resistance of different heating schemes}
\label{sec: TR results}
With the help of the hybrid method, it is possible to conduct in detail studies about the thermal spreading resistance, including the cases with relatively large system thicknesses. The total thermal spreading resistance is calculated using the average temperature rise of the heat generation region \cite{MuzychkaCulham-828}:
\begin{equation}
    R=\frac{\int_{\rm heating\ region}(T-T_0){\rm d}V}{\dot{Q}_{\rm s} L_{\rm s} t_{\rm s}}
\end{equation} 
The current problem involves four factors that may jointly determine the thermal spreading resistance: (1) the thermal spreading effect that is associated with the system shape; (2) the cross-plane ballistic effect that is controlled by the system thickness; (3) the lateral ballistic effect that occurs because the width of the heat generation region is comparable with the phonon MFP; and (4) the ballistic effect that depends on the thickness of the heat generation region. Our previous work \cite{HuaLi-661} investigated the effects of the first three factors in the HF scheme and established a semi-empirical thermal spreading resistance model as $R_{\rm HF}=R_{\rm F}(1+\frac{2}{3}Kn_t)(1+A_L Kn_{L,{\rm s}})$, which is dependent on  $L/t$, $L_{\rm s}/L$ and $Kn_t$. In the HS scheme, when the thickness of the heat generation region, $t_{\rm s}$, is comparable to the phonon MFP, the fourth factor could produce a new kind of ballistic effect and affect the thermal spreading resistance. Since our attention is given to the dependence of phonon transport on $Kn_{t,{\rm s}}$, all the calculations in this section are conducted under the same values of $L_{\rm s}/L=0.01$ and $L/t=40$. As mentioned in Sec. \ref{sec: validation}, for the cases of relatively large system thicknesses ($Kn_t \le 0.2$), the hybrid MC-diffusion method is utilized; for the cases of relatively small system thicknesses ($Kn_t > 0.2$), MC simulation is a preferable choice.  

\begin{figure}
    \centering
    \begin{minipage}{80mm}
        \centering
        \includegraphics[width=1.0\linewidth,height=78mm]{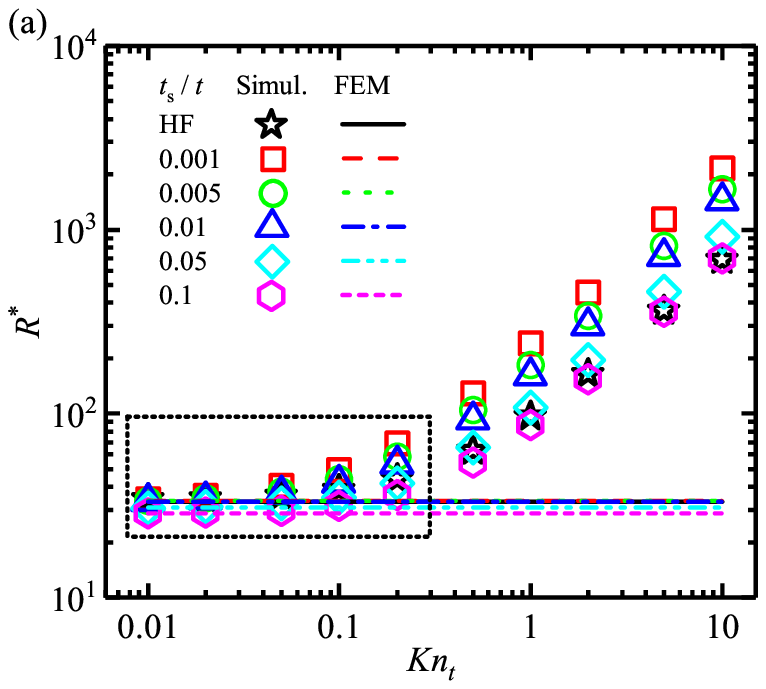}
    \end{minipage}
    \begin{minipage}{80mm}
        \centering
        \includegraphics[width=1.0\linewidth,height=78mm]{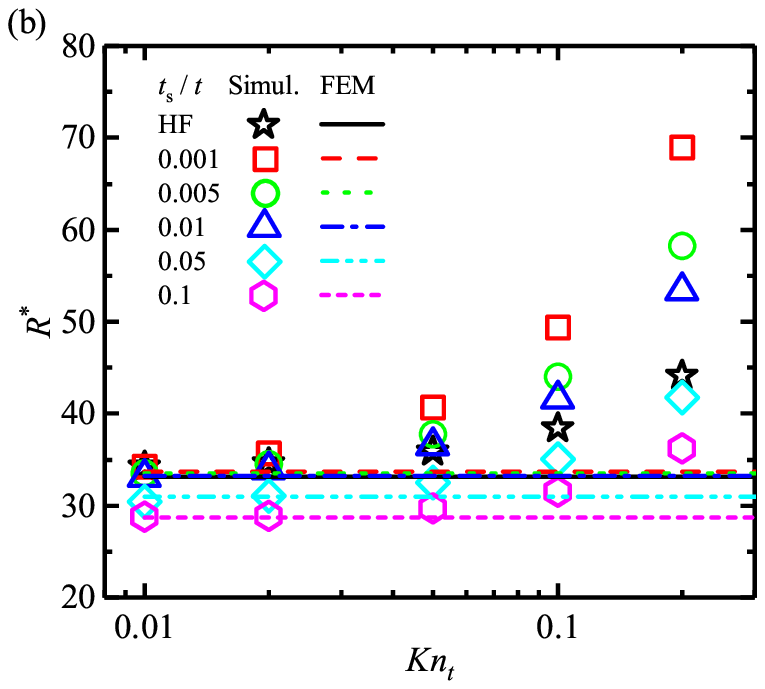}
    \end{minipage}
    \caption{Dimensionless thermal spreading resistance varying with $Kn_t$: (a) $0.01 \le Kn_t \le 10$; (b) enlarged at $0.01 \le Kn_t \le 0.2$. ($L_{\rm s}/L=0.01$, $L/t=40$)}
    \label{fig: TR_star}
\end{figure}

The dimensionless thermal spreading resistance, $R^*=R/R_{\rm 1D,0}$, varying with the system thickness is shown in Fig. \ref{fig: TR_star} (a), where the results for $Kn_t \le 0.2$ are more distinctly demonstrated in Fig. \ref{fig: TR_star} (b). It is found that the heating scheme indeed plays an important role in the thermal spreading resistance, and the influence is closely related to $Kn_t$. The dimensionless thermal resistance based on Fourier's law, $R^*_{\rm FEM}$, is determined by the system shape, which is controlled by $L_{\rm s}/L$, $L/t$ and $t_{\rm s}/t$, so it will not vary with $Kn_t$, as the constant lines in Fig. \ref{fig: TR_star} exhibit. When the heating scheme changes, $R^*_{\rm FEM}$ varies with the heat source thickness. For $t_{\rm s}/t \le 0.01$, the dimensionless thermal resistance in the HS scheme is $R^*_{\rm FEM,HS} \approx 33.5$, which is almost the same as the value of the HF scheme ($R^*_{\rm FEM,HF} = 33.1$). With $t_{\rm s}/t$ rising, $R^*_{\rm FEM,HS}$ declines. For example, $R^*_{\rm FEM,HS} = 28.7$ at $t_{\rm s}/t = 0.1$ is about $10\%$ less than $R^*_{\rm FEM,HF}$. The $t_{\rm s}/t$-dependent thermal spreading resistance under Fourier's law is attributed to the heat accumulation effect. As $t_{\rm s}/t$ decreases, heat dissipation is constrained to a thinner region, and more accumulated heat produces a higher junction temperature and larger thermal spreading resistance. When $t_{\rm s}/t$ is small enough, say $t_{\rm s}/t \le 0.01$, the heat generation region can be approximately viewed as a ``line". As a result, $R^*_{\rm FEM,HS}$ approaches $R^*_{\rm FEM,HF}$.

In ballistic-diffusive regime, ballistic effect amplifies the thermal spreading resistance; thus, the symbols in Fig. \ref{fig: TR_star} lie above the lines. Considering the $t_{\rm s}/t$ dependence of $R^*$ in Fourier's law, $Kn_{t,{\rm s}}$ is not explicitly exhibited because $Kn_{t,{\rm s}}=\frac{Kn_t}{t_{\rm s}/t}$. For $Kn_t=0.01$, the thickness is so large that most phonons are transported diffusively, and $R^*$ of the hybrid MC-diffusion method compares favorably to the FEM solutions. As $Kn_t$ advances, phonon ballistic transport appears and the dimensionless thermal spreading resistances are greater than the results of FEM, but the strength of ballistic effect is associated with $t_{\rm s}/t$. The thinner the heat generation region is, the earlier the results of the hybrid MC-diffusion method depart from those of FEM. As presented in Fig. \ref{fig: TR_star} (b), for $t_{\rm s}/t=0.001$, the hybrid method exceeds the FEM solution at $Kn_t=0.02$, but for $t_{\rm s}/t=0.1$, the departure takes place at $Kn_t=0.05$. Furthermore, the difference between the HS scheme and HF scheme also differs as $Kn_t$ alters, especially in the range of $Kn_t \le 0.2$. For $Kn_t=0.01$, $R^*_{\rm HS}$ with $t_{\rm s}/t \le 0.01$ is almost the same as $R^*_{\rm HF}$. For $Kn_t=0.02$, $R^*_{\rm HS}$ with $t_{\rm s}/t = 0.001$ is slightly larger than $R^*_{\rm HF}$. When $Kn_t$ increases to 0.05, $R^*_{\rm HS}$ with $t_{\rm s}/t = 0.005$ also surpasses $R^*_{\rm HF}$. For $Kn_t=0.1$, the transcendence of $R^*_{\rm HS}$ with $t_{\rm s}/t = 0.01$ over $R^*_{\rm HF}$ emerges. For $Kn_t=0.2$, $R^*_{\rm HF}$ only corresponds to $R^*_{\rm HS}$ with $t_{\rm s}/t = 0.05$. As $Kn_t$ continues to increase, the profile of $R^*_{\rm HF}$ gets lower. Fig. \ref{fig: TR_star} (a) displays that $R^*_{\rm HS}$ with $t_{\rm s}/t = 0.1$ is comparable to $R^*_{\rm HF}$ after $Kn_t>1$. It is illustrative that $R^*_{\rm HF}$ is no longer the upper bound on $R^*_{\rm HS}$ as the ballistic effect strengthens, which is in contrast with what is observed under Fourier's law.

\begin{figure}
    \centering
    \begin{minipage}{80mm}
        \centering
        \includegraphics[width=1.0\linewidth,height=78mm]{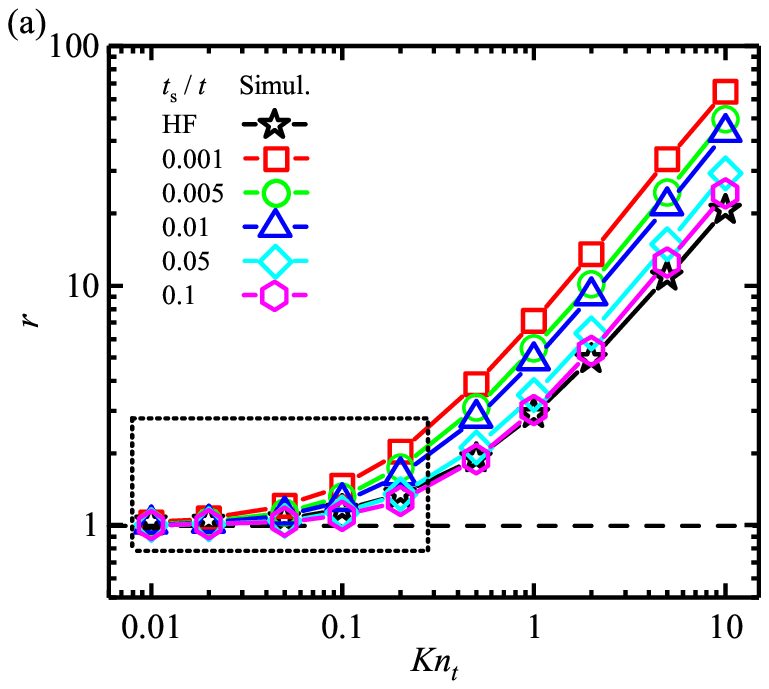}
    \end{minipage}
    \begin{minipage}{80mm}
        \centering
        \includegraphics[width=1.0\linewidth,height=78mm]{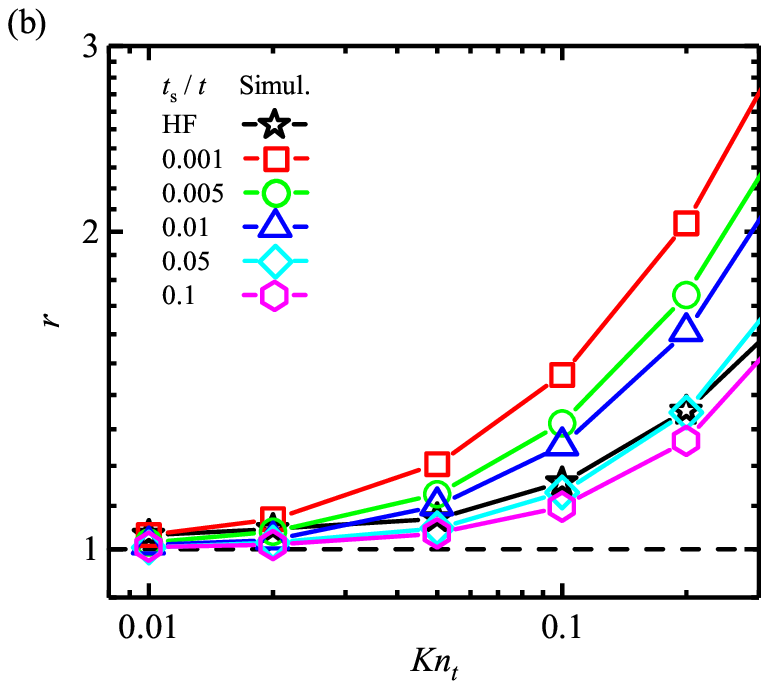}
    \end{minipage}
    \caption{Thermal spreading resistance ratio varying with $Kn_t$: (a) $0.01 \le Kn_t \le 10$; (b) enlarged at $0.01 \le Kn_t \le 0.2$. ($L_{\rm s}/L=0.01$, $L/t=40$)}
    \label{fig: TR ratio}
\end{figure}

In order to clarify the influence of the heating scheme on phonon ballistic transport, the thermal spreading resistance ratio, $r=R_{\rm Hybrid/MC}/R_{\rm FEM}$, is introduced, and the results are illustrated in Fig. \ref{fig: TR ratio}. Here, $R_{\rm Hybrid/MC}$ denotes the thermal spreading resistance calculated by the hybrid method (for $Kn_t \le 0.2$) or MC simulation (for $Kn_t > 0.2$). All the results are larger than 1, and the values move upward as $Kn_t$ increases, indicating the appearance of phonon ballistic transport. It is emphasized that even for relatively large system thicknesses ($Kn_t \le 0.2$), ballistic effect cannot be ignored, as $r$ in Fig. \ref{fig: TR ratio} (b) is definitely greater than 1. Taking the HS scheme with $t_{\rm s}/t=0.001$ as an example, $r \approx 1.2$ at $Kn_t = 0.05$ means that using FEM produces an error of $20\%$ in this case. Fig. \ref{fig: TR ratio} also reveals that the heating scheme adjusts the strength of ballistic effect. In the HS scheme, cutting down $t_{\rm s}/t$ raises $r$, and the $r$ profile of the HF scheme gradually becomes lower than the results of the HS scheme with ballistic effect enhancement. For $Kn_t = 0.01$, $r_{\rm HF}$ approximately equals $r_{\rm HS}$ with $t_{\rm s}/t \le 0.01$. For $Kn_t = 0.05$, $r_{\rm HF}$ is between the values of $r_{\rm HS}$ with $t_{\rm s}/t=0.01$ and $r_{\rm HS}$ with $t_{\rm s}/t=0.05$. For $Kn_t>1$, even $r_{\rm HS}$ with $t_{\rm s}/t=0.1$ is kind of higher than $r_{\rm HF}$. Fig. \ref{fig: TR ratio} proves that reducing the thickness of the heat source yields a stronger ballistic effect, making it possible to overtake the strength of ballistic effect in the HF scheme. 

\begin{figure}[htbp]
	\centering
    \includegraphics[width=80mm,height=78mm]{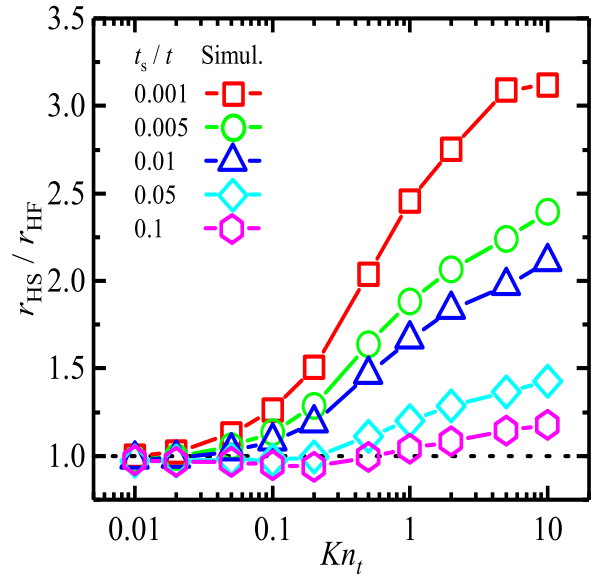}
    \caption{Ratio of $r_{\rm HS}$ to $r_{\rm HF}$ varying with $Kn_t$ ($L_{\rm s}/L=0.01$, $L/t=40$)}
    \label{fig: r ratio of HS to HF 40}
\end{figure}

To more directly compare ballistic effect in the HS scheme and the HF scheme, we calculate the ratio of $r_{\rm HS}$ to $r_{\rm HF}$ and get the results displayed in Fig. \ref{fig: r ratio of HS to HF 40}. Starting from 1, the values of $r_{\rm HS}/r_{\rm HF}$ increase with increasing $Kn_t$, demonstrating again that the HS scheme is able to generate a stronger ballistic effect compared to the HF scheme. A thinner heat generation region will accelerate this process of enlargement. For instance, with $t_{\rm s}/t=0.001$, $r_{\rm HS}/r_{\rm HF}$ has been evidently greater than 1 at $Kn_t=0.1$; however, with $t_{\rm s}/t=0.1$, the excess of $r_{\rm HS}/r_{\rm HF}$ on 1 does not turn up until $Kn_t$ is beyond 1. There are even some places where $r_{\rm HS}/r_{\rm HF} < 1$, say, $t_{\rm s}/t=0.1$ and $Kn_t=0.2$. 

\subsection{Analyses of the effects of the heating scheme}
\label{sec: Analysis}
According to the results in Sec. \ref{sec: TR results}, it is fair to conclude that the heating scheme does have a powerful impact on the thermal spreading resistance, and the HS scheme with a thinner heat generation region induces stronger phonon ballistic transport than the HF scheme. The findings can be explained as the change in the relative intensity of phonon-boundary scattering at the top boundary. In the HF scheme, heat is generated at the top heat flux boundary, all the motivated phonons move downward, and they have no chance of suffering the scattering of the top boundary unless intrinsic phonon-phonon scattering takes place. However, in the HS scheme, heat is generated inside the system, and the initial phonon movement direction is spatially isotropic. Statistically speaking, half of the phonons go upward after emitting. Since the heat generation region is near the top boundary, almost all upward-moving phonons travel to the top boundary in a ballistic way and suffer phonon-boundary scattering instead of phonon-phonon scattering, which sharply curtails their effective MFPs. As a consequence, the effective thermal conductivity is reduced \cite{CaoLi-507,LiCao-506,HuaCao-70}, and the thermal resistance is improved. A greater $Kn_{t,{\rm s}}$ is supposed to reinforce this kind of heat-source-thickness-related ballistic effect. When the heating scheme changes from HF to HS, on the one hand, the transform of a line heat source to a rectangle heat source diminishes the heat accumulation effect, which is helpful to decrease the thermal spreading resistance; on the other hand, the enhanced scattering at the top boundary provides more boundary confined effects on phonon transport, which raises the thermal spreading resistance. The difference between $r_{\rm HS}$ and $r_{\rm HF}$ is a balance of the two mechanisms. For $Kn_t=0.01$, phonon-phonon scattering is sufficient to mask the influence of phonon-boundary scattering, and the heat generation region thickness has little effect on the thermal spreading resistance, so $r_{\rm HS} / r_{\rm HF} \approx 1$. As $Kn_t$ increases, the proportion of phonon-boundary scattering increases, and the $Kn_{t,{\rm s}}$-dependent phonon ballistic transport begins to take effect, resulting in $r_{\rm HS} / r_{\rm HF} > 1$. A small heat generation thickness such as $t_{\rm s}/t=0.001$ boosts $Kn_{t,{\rm s}}$ to be much larger than 1, and $r_{\rm HS} / r_{\rm HF}>1$ has been visible after $Kn_t \ge 0.05$ in Fig. \ref{fig: r ratio of HS to HF 40}. Raising $t_{\rm s}/t$ weakens the $Kn_{t,{\rm s}}$-dependent phonon ballistic transport and delays the point where $r_{\rm HS} / r_{\rm HF}>1$ happens. In some particular cases where the thermal resistance reduction caused by the thicker heat source covers the enlargement by the strengthened phonon-boundary scattering, $r_{\rm HS} / r_{\rm HF}$ could be less than 1, as the result of $t_{\rm s}/t=0.1$ and $Kn_t=0.2$ in Fig. \ref{fig: r ratio of HS to HF 40} exhibits.

To test the validity of the analysis, a model about $r_{\rm HS} / r_{\rm HF}$ is established. For the purpose of focusing on the $Kn_{t,{\rm s}}$-dependent ballistic effect, $L_{\rm s}/L$ is set to 1 to avoid the interference of the thermal spreading effect and heat-source-width-dependent ballistic effect. Under this condition, the problem degenerates to 1D heat conduction with a small heating region, and phonon BTE can be analytically solved with some approximations. Considering the heat is conducted in the $z$ direction, the 1D form of phonon BTE is written as
\begin{subequations}
    \begin{align}
        v_{{\rm g},z}\frac{\partial f}{\partial z} & = \frac{f_0-f}{\tau}+\dot{S}_{\Omega}, 0\le z \le t_{\rm s}\\
        v_{{\rm g},z}\frac{\partial f}{\partial z} & = \frac{f_0-f}{\tau}, t_{\rm s} \le z \le t
    \end{align}
    \label{eq: BTE in 1D}
\end{subequations}
Here, $v_{{\rm g},z}=v_{\rm g}\cos\theta$in which $\theta$ is the angle between the phonon traveling direction and the $z$ direction, and $v_{\rm g}$, $f$, $f_0$, $\tau$ and $\dot{S}_{\Omega}$ denote the phonon group velocity, phonon distribution function, equilibrium distribution function, relaxation time and phonon source per solid angle, respectively. To solve Eq. (\ref{eq: BTE in 1D}), the distribution function is divided into two parts, $f=f_{\rm s}+f_{\rm d}$ \cite{HuaCao-70}, in which $f_{\rm s}$ is the source-induced part that will be solved by the two-flux approximation and $f_{\rm d}$ is the diffusive part that will be solved by the differential approximation. After a series of mathematical derivations, we have
\begin{equation}
    r_{\rm 1D,HS}  =\frac{R_{{\rm 1D,HS}}}{R_{\rm 1D,HS,FEM}}=1+\frac{\frac{2}{3}+b}{1-\frac{2}{3}a}Kn_t
    =1+\alpha \cdot Kn_t
    \label{eq: r_1DHS}
\end{equation}
in which $a =t_{\rm s}/t$, $b=\frac{Kn_t}{24a}[\exp(-2\frac{1+a}{Kn_t})-\exp(-2\frac{1-a}{Kn_t})+2]+\frac{Kn_t^2}{48a^2}[\exp(-4\frac{a}{Kn_t})-1]$. 
Equation (\ref{eq: r_1DHS}) indicates that the thermal resistance ratio can be expressed as a function of the shape factor $\alpha$ and the Knudsen number $Kn_t$ \cite{HuaCao-67}. For the HS scheme, it is reported that $r_{\rm 1D,HF}=1+\frac{2}{3}Kn_t$ \cite{HuaLi-661}. Thus, we have
\begin{equation}
    r_{\rm 1D,HS}/r_{\rm 1D,HF}=\frac{1+\alpha Kn_t}{1+\frac{2}{3}Kn_t}
    \label{eq: r_1DHS to r_1DHF}
\end{equation}
Equation (\ref{eq: r_1DHS to r_1DHF}) implies that the magnitude relationship between $\alpha$ and $\frac{2}{3}$ decides the value of $r_{\rm 1D,HS}/r_{\rm 1D,HF}$. Within the range of $t_{\rm s}/t \in [0.001,0.1]$, it is always true that $b>0$ and $\alpha > \frac{2}{3}$. Consequently, the effect of phonon ballistic transport in the HS scheme is more conspicuous than that in the HF scheme. 

\begin{figure}
    \centering
    \begin{minipage}{80mm}
        \centering
        \includegraphics[width=1.0\linewidth]{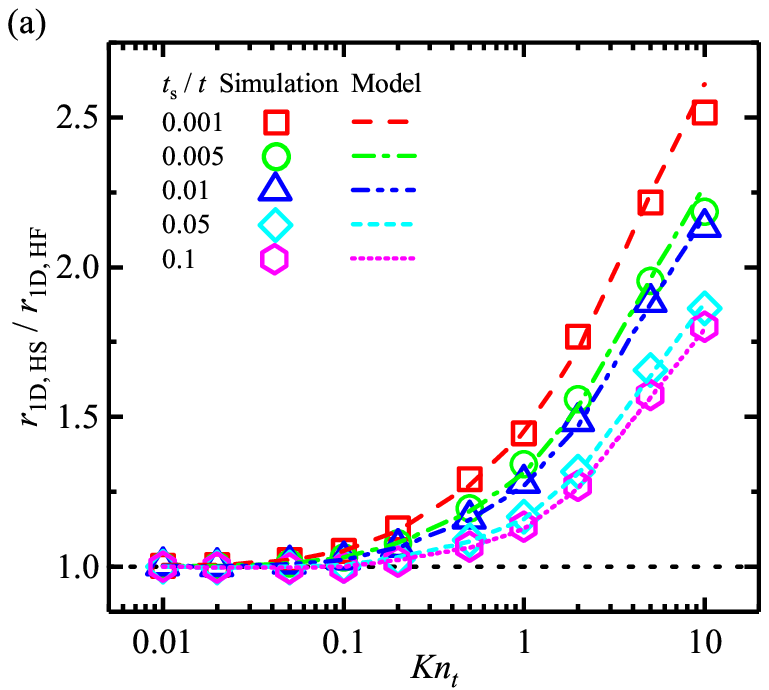}
    \end{minipage}
    \begin{minipage}{80mm}
        \centering
        \includegraphics[width=1.0\linewidth]{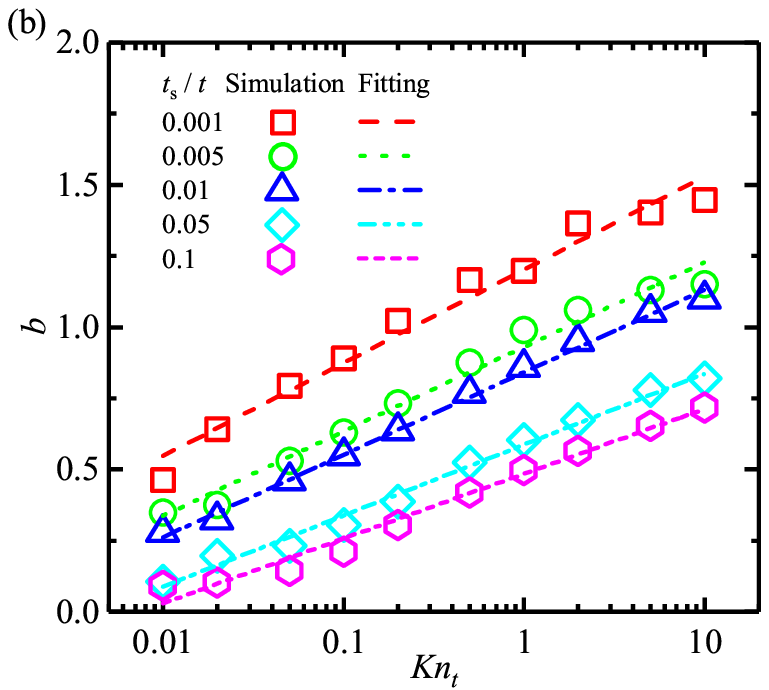}
    \end{minipage}
    \caption{(a) Ratio of $r_{\rm HS}$ to $r_{\rm HF}$ varying with $Kn_t$ in 1D cases; (b) the modified values of $b$.}
    \label{fig: 1D r ratio}
\end{figure}

The results of $r_{\rm HS}/r_{\rm HF}$ in 1D cases calculated by the hybrid MC-diffusion method or MC simulation are depicted in Fig. \ref{fig: 1D r ratio} (a). The original results of Eq. (\ref{eq: r_1DHS to r_1DHF}) have some quantitative differences with the simulation results, which is probably proposed by the approximations used in the derivation, like dividing the distribution function, selecting the boundary condition for ballistic-diffusive heat conduction and adopting the differential approximation. To modify the model, we calculate the specific values of $b$ in Eq. (\ref{eq: r_1DHS}) by comparing with the simulation, and the obtained results are drawn in Fig. \ref{fig: 1D r ratio} (b). $b$ is found to increase with $Kn_t$, and a smaller $t_{\rm s}/t$ leads to a larger $b$. Moreover, the linear fitting of $\ln(Kn_t)$ agrees well with the simulation results, so that $b$ can be predicted by the fitting functions. Using the fitted values of $b$, the current model does a reasonably good job of matching the simulation, as the lines in Fig. \ref{fig: 1D r ratio} (a) shows.

Although $b$ needs to be modified quantitatively, the expression that $\alpha=\frac{\frac{2}{3}+b}{1-\frac{2}{3}a}$ helps to understand the difference between the HS scheme and the HF scheme. The denominator ``$1-\frac{2}{3}a$'' reflects the variation of heat source thickness; a greater $a=t_{\rm s}/t$ means heat is generated in a wider region, the deviation from a ``line'' heat source is more apparent and $\alpha$ is larger. The numerator ``$\frac{2}{3}+b$'' reflects the combination of two kinds of phonon ballistic transport, involving the cross-plane ballistic transport represented by ``$\frac{2}{3}$'' and the heat-source-thickness-dependent ballistic transport represented by ``$b$'' ($b$ is explicitly dependent on $Kn_{t,{\rm s}}=Kn_t/a$). The division operation uncovers that there is a trade-off between the heat accumulation effect and the $Kn_{t,{\rm s}}$-related ballistic effect. In the limit case of no thickness heat source ($a=0$) and no heat-source-thickness-dependent ballistic effect ($b=0$), Eq. (\ref{eq: r_1DHS}) gives $\alpha=\frac{2}{3}$, reproducing the results of the HF scheme. 

\section{Conclusion}
In the present work, a hybrid phonon MC-diffusion method is applied to thermal spreading problems in ballistic-diffusive heat conduction. By setting the MC zones that cover the heat generation region and the boundaries, the hybrid MC-diffusion method captures the elevated junction temperature and the recognizable boundary temperature jump that cannot be observed by Fourier's law. Compared with MC simulation, the hybrid method reduces the computational time up to 2 orders of magnitude at most, while the temperature relative error is less than $5\%$, which facilitates the accurate and fast simulation of the thermal spreading process in relatively large systems ($Kn_t \le 0.2$).

The simulation results suggest that the thermal spreading resistance in ballistic-diffusive regime is certainly sensitive to the heating scheme: the HS scheme corresponds to a higher junction temperature and thermal spreading resistance than the HF scheme as long as $Kn_t$ is large enough, and a thinner heat generation region thickness reinforces this deviation. For $Kn_t \le 0.2$, ballistic effect in the HF scheme is roughly equivalent to that in the HS scheme with $t_{\rm t}/t=0.05$; for the greater $Kn_t$, the HS scheme even with $t_{\rm t}/t=0.1$ can yield stronger ballistic effect than the HF scheme. The simulation findings are supported by an analytical model based on the 1D phonon BTE. 

The dependence of thermal spreading resistance on the heating scheme is attributed to the relative strength of phonon-boundary scattering in phonon transport. In the diffusive regime, phonon-phonon scattering dominates heat transfer, and the thermal resistance in the HF scheme works as the upper bound of that in the HS scheme since it has the thinnest heat source. However, in the cases where ballistic effect cannot be ignored, phonon boundary scattering at the top boundary in the HS scheme is more intense than in the HS scheme and strengthens with $t_{\rm t}/t$, which results in stronger ballistic effect as well as larger thermal spreading resistance. To evaluate the thermal spreading resistance precisely, specific analysis must be made for different geometry sizes and heating schemes.

\section*{Acknowledgement}
This work is financially supported by National Natural Science Foundation of China (Nos. U20A20301, 51825601).
The reported study was also funded by Russian Foundation for Basic Research and National Natural Science Foundation of China (Nos. 20-58-53017, 52011530030).

\bibliography{./refs}

\newpage
\listoffigures

\end{document}